\documentclass[fleqn,usenatbib]{mnras}

\usepackage{newtxtext,newtxmath}

\usepackage[T1]{fontenc}
\usepackage{booktabs}
\usepackage{amsfonts}

\DeclareRobustCommand{\VAN}[3]{#2}
\let\VANthebibliography\thebibliography
\def\thebibliography{\DeclareRobustCommand{\VAN}[3]{##3}\VANthebibliography}

\usepackage{graphicx}	
\usepackage{subcaption}
\usepackage{amsmath}	

\title[Test CDDR using NKGPR]{Testing the Cosmic Distance Duality Relation with Neural Kernel Gaussian Process Regression}

\author[Luo and Liang]{
Xin Luo$^{1,2}$\thanks{E-mail: \href{mailto:xin_luo@gznu.edu.cn}{xin\_luo@gznu.edu.cn}}
Nan Liang,$^{1}$\thanks{E-mail: \href{mailto:liangn@bnu.edu.cn}{liangn@bnu.edu.cn}}
\\
$^{1}$Guizhou Key Laboratory of Advanced Computing, Guizhou Normal University, Guiyang, Guizhou 550025, China\\
$^{2}$School of Cyber Science and Technology, Guizhou Normal University, Guiyang, Guizhou 550025, China\\
}

\date{Accepted XXX. Received YYY; in original form ZZZ}

\pubyear{\the\year{}}

\begin{document}
\label{firstpage}
\pagerange{\pageref{firstpage}--\pageref{lastpage}}
\maketitle

\begin{abstract}

In this work, we test the cosmic distance duality relation (CDDR) by combining Pantheon+ Type Ia supernova (SNe Ia) data and DESI DR2 baryon acoustic oscillation (BAO) measurements.
To resolve the redshift mismatch between the two datasets, we develop a new method called Neural Kernel Gaussian Process Regression (NKGPR), which uses two neural networks to simultaneously learn the mean and kernel functions of a Gaussian process.
This approach improves upon traditional Gaussian process regression by mitigating trend mismatches and removing the need for manual kernel selection.
We investigate possible deviations from the CDDR by adopting three parameterizations of the deviation function and constrain the model-independent parameter $\eta_0$ through a marginalized likelihood analysis.
Our results show no significant departure from the expected relation, confirming the consistency of the CDDR within current observational uncertainties.

\end{abstract}

\begin{keywords}
Cosmology - software: machine learning - galaxies: distances and redshifts - methods: data analysis
\end{keywords}



\section{Introduction}

The cosmic distance duality relation (CDDR), first proved by \citet{Etherington1933}, describes a fundamental relationship between the luminosity distance (LD) $D_{\mathrm{L}}$ and the angular diameter distance (ADD) $D_{\mathrm{A}}$ as a function of redshift ($z$):
\begin{equation}
	\frac{D_{\mathrm{L}}}{D_{\mathrm{A}}} (1+z)^{-2} = 1.
\end{equation}
It holds in cosmological models based on Riemannian geometry, independent of the specific metric or cosmological framework, relying on two fundamental assumptions: (a) photons propagate along null geodesics in a Riemannian spacetime; (b) the number of photons is conserved throughout cosmic evolution. The CDDR provides an important test of cosmological models and offers insights into the nature of cosmic expansion and dark energy. Any observed deviation from the CDDR may suggest either new physical phenomena or unaccounted observational systematics \citep{Bassett2004, Holanda2010}.

Extensive efforts have been made to test the CDDR using various astronomical observations.
Type Ia supernovae (SNe Ia) are commonly used to determine LD due to their well-established role as standard candles \citep{Riess1998}. Meanwhile, ADD is typically obtained via multiple observational techniques, including the Sunyaev-Zeldovich effect \citep{Ameglio2006,Holanda2019}, gas mass fraction measurements in galaxy clusters \citep{Gonçalves2012}, baryon acoustic oscillations (BAOs) \citep{Ma2018,Wang2024}, and strong gravitational lensing (SGL) systems \citep{Liao2016,Tang2024}. These studies did not find significant deviations from CDDR. However, recent results from the DESI Collaboration 
\citep{DESI:2404.03002, DESI:2411.12022, DESI:2503.14738, DESI:2503.14739}
seem to suggest that a dynamical dark energy scenario may be favored over a pure cosmological constant \citep{lodha2025,rodrigues2025}. Motivated by this, we adopt the latest DESI BAO data to obtain the ADD measurements.

One challenge in testing the CDDR arises from the redshift mismatch between LD measurements from SNe Ia and ADD estimates from various observations. Earlier studies addressed this issue by selecting SNe Ia whose redshifts are closest to the galaxy cluster's within a small interval $\Delta z=|z-z_{\rm SNe Ia}|<0.005$ \citep{Holanda2010,Li2011}. \citet{Meng2012} extended this method by binning SNe Ia data within $\Delta z$. 
To overcome redshift mismatch biases and preserve the integrity of galaxy cluster samples, \cite{Liang2013} introduced a new consistent method to test the CDDR. This method derives the LD of a certain SN Ia point at the same redshift of the corresponding galaxy cluster by interpolating from the nearby SNe Ia, effectively relating LD to ADD without cosmological assumptions, which is similar with  the cosmology-independent calibration of GRB relations directly from SNe Ia \citep{Liang2008, Liang2010, Liang2011}.
\citet{Chen2016} tested the CDDR by reconstructing the LD by smoothing the noise of Union2.1 data set over redshift with the binning method and the Gaussian smoothing function.

To handle the correlations in astronomical observational data, \citet{Seikel2012} applied Gaussian Process Regression (GPR) to the reconstruction of $H(z)$ data. Since then, GPR has been widely used in various fields of astronomy \citep{Zhou2019,Mehrabi2021,Liang2022,Li2023,WangL2024,Zhang2024}.
Recently, many works \citep{Lin2018, Mukherjee2021, Wang2024} have tested the possible violation of CDDR by using GPR, with the available local data including SNe Ia, galaxy clusters, the cosmic chronometer (CC) Hubble parameter, and BAO.
However, the results of GPR become unreliable with large uncertainties at high redshift ($z>1$) where data are sparse \citep{Zhou2019,Lin2018}. Moreover, the choice of kernel function influences the reconstruction results, and selecting an appropriate kernel often relies on subjective experience \citep{Bernardo2021,Zhang2023,johnson2025}.

With the development of deep learning techniques, researchers have explored their applications in astronomy to tackle complex data challenges. 
\citet{Wang2020}
proposed a non-parametric approach using artificial neural networks (ANNs) to reconstruct functions from observational data. 
More recently, several studies have applied ANNs to test the CDDR \citep{Liu2021, Xu2022, Tang2023}. \citet{Yang2025} proposed to utilize the observed ratio from BAO data by comparing the LD obtained from SNIa observations with ADD with ANN method to test the CDDR. 
\cite{Huang2025} employed an ANN framework to reconstructe $H(z)$ in a model-independent way by considering the physical correlations in the data with the covariance matrix and KL (Kullback-Leibler) divergence into the loss function.
ANNs can automatically learn complex relationships from observational data without relying on cosmological model assumptions. However, the ANN method also has the following drawbacks: it is difficult to directly account for the correlations among observational data; and the interpretability of the generated errors is not as good as that of GPR.

To integrate the respective strengths of ANNs and GPR, \citet{Wilson2016} introduced deep kernel learning (DKL). In this framework, a neural network maps the input data into a high-dimensional feature space, enhancing data separability and enabling GPR to capture more complex patterns. Despite its effectiveness, DKL remains dependent on predefined kernel functions, such as the radial basis function (RBF), which can limit its flexibility in modeling specific dataset correlations.

Motivated by these limitations, we propose a new method, termed Neural Kernel Gaussian Process Regression (NKGPR), to reconstruct the SNe Ia dataset. In contrast to conventional GPR frameworks that require manually defined mean and kernel functions, NKGPR employs two dedicated ANNs to learn these components directly from data. One network models the mean function, while the other learns the kernel structure.
To guarantee symmetry and positive semi-definiteness in the learned kernel, the kernel network is structured to output a lower triangular matrix, which is then multiplied by its transpose to produce the final covariance matrix. 
Bayesian inference is subsequently performed using the learned mean and covariance functions. Unlike DKL, which indirectly influences the kernel through feature transformations, NKGPR directly models the covariance matrix, thereby eliminating the need for handcrafted kernels. This direct learning framework provides greater adaptability to the sparse, noisy, and non-stationary nature of high-redshift cosmological data.

In this paper, we use the NKGPR method to test the CDDR with three parameterized forms to account for possible violations, P1: $\eta(z)=1+\eta_0z$, P2: $\eta(z)=1+\eta_0z/(1+z)$, and P3: $\eta(z)=1+\eta_0\ln(1+z)$. Any nonzero $\eta_0$ indicates a deviation of the CDDR from observational data. Furthermore, to mitigate the influence of prior assumptions on $M_{\mathrm{B}}$ and $r_{\rm d}$, we follow the approach proposed by \citet{Wang2024}, treating them as nuisance parameters and marginalizing over them.

This paper is organized as follows. Section~\ref{sec:DATA AND METHODOLOGY} introduces the data and methodology used in our analysis. Section~\ref{sec:RESULTS AND ANALYSIS} presents the results and discusses their implications. Finally, Section~\ref{sec:CONCLUSION} summarizes our conclusions.

\section[DATA AND METHODOLOGY]{DATA AND METHODOLOGY}\label{sec:DATA AND METHODOLOGY}

\subsection{Data}

In this study, we use the Pantheon+ SNe Ia sample \citep{Scolnic2022} to compute LD and the latest BAO data: DESI DR2
\citep{DESI:2503.14738,DESI:2503.14739}
to derive ADD.

\subsubsection{Pantheon+ Supernova Sample}
The Pantheon+ sample is the latest compilation of SNe Ia used for cosmological distance measurements. It consists of 1701 light curves from 1550 SNe Ia over a redshift range of $0.001<z<2.26$, offering a high-precision constraint on the expansion history of the universe.

The key observational quantities provided by Pantheon+ include the apparent B-band magnitude $m_{\mathrm{B}}$ and redshift $z$. The LD can be related to the distance modulus $\mu$ via the standard formula:
\begin{equation}
	\mu = m_{\mathrm{B}} - M_{\mathrm{B}} = 5 \log_{10}\left(\frac{D_{\mathrm{L}}}{\text{Mpc}}\right) + 25,
\end{equation}
where $M_{\mathrm{B}}$ is the absolute magnitude of SNe Ia. By rearranging, we obtain:
\begin{equation}
	D_{\mathrm{L}} = 10^{\frac{\mu - 25}{5}}\quad\text{Mpc}.
\end{equation}

\subsubsection{DESI BAO Data}
The Dark Energy Spectroscopic Instrument (DESI) provides the latest measurements of BAO, offering a robust probe of cosmic expansion. The latest BAO results: DESI DR2
\citep{DESI:2503.14738,DESI:2503.14739}
report measurements of the dimensionless ratio $D_{\mathrm{M}}/r_{\mathrm{d}}$, where $D_{\mathrm{M}}$ is the comoving angular diameter distance and $r_{\mathrm{d}}$ is the sound horizon at the drag epoch.

The ADD can be obtained using:
\begin{equation}
	D_{\mathrm{A}} = \frac{D_{\mathrm{M}}}{1+z}.
\end{equation}
These ADD measurements, derived from DESI BAO data, are combined with Pantheon+ SNe Ia LD to test the CDDR, with $r_{\mathrm{d}}$ and $M_{\mathrm{B}}$ treated as nuisance parameters and marginalized over. The set of DESI BAO measurements used in this study is presented in Table~\ref{tab:desi_bao_full}.

\begin{table}
	\centering
	\caption{DESI DR2 BAO measurements from \citet{DESI:2503.14738,DESI:2503.14739}, providing $D_{\mathrm{M}}/r_{\mathrm{d}}$, $D_{\mathrm{H}}/r_{\mathrm{d}}$, and $D_{\mathrm{V}}/r_{\mathrm{d}}$ at various effective redshifts $z_{\text{eff}}$. In this study, we use the six tracers with available $D_{\mathrm{M}}/r_{\mathrm{d}}$ constraints (LRG1, LRG2, LRG3+ELG1, ELG2, QSO and Lya) for the CDDR test.}
	\label{tab:desi_bao_full}
	\begin{tabular}{l@{\hspace{6pt}}c@{\hspace{6pt}}c@{\hspace{6pt}}c@{\hspace{6pt}}c}
		\toprule
		Tracer & $z_{\text{eff}}$ & $D_{\mathrm{V}}/r_{\mathrm{d}}$ & $D_{\mathrm{M}}/r_{\mathrm{d}}$ & $D_{\mathrm{H}}/r_{\mathrm{d}}$ \\
		\midrule
		BGS       & 0.295 & 7.942 $\pm$ 0.075 & —              & —              \\
		LRG1      & 0.510 & 12.720 $\pm$ 0.099 & 13.588 $\pm$ 0.167 & 21.863 $\pm$ 0.425 \\
		LRG2      & 0.706 & 16.050 $\pm$ 0.110 & 17.351 $\pm$ 0.177 & 19.455 $\pm$ 0.330 \\
		LRG3+ELG1      & 0.934 & 19.721 $\pm$ 0.091 & 21.576 $\pm$ 0.152 & 17.641 $\pm$ 0.193 \\
		ELG2      & 1.321 & 24.252 $\pm$ 0.174 & 27.601 $\pm$ 0.318 & 14.176 $\pm$ 0.221 \\
		QSO       & 1.484 & 26.055 $\pm$ 0.398 & 30.512 $\pm$ 0.760 & 12.817 $\pm$ 0.516 \\
		Lya       & 2.330 & 31.267 $\pm$ 0.256 & 38.988 $\pm$ 0.531 & 8.632 $\pm$ 0.101 \\
		\bottomrule
	\end{tabular}
\end{table}

\subsection{METHODOLOGY}

\subsubsection{Gaussian Process Regression}

GPR is a non-parametric Bayesian method used to reconstruct a continuous function from observed data. The key idea is to assume that the target function $f(x)$ at any input $x$ follows a Gaussian process, meaning that the function values at any finite set of inputs are jointly distributed as a multivariate normal distribution \citep{Rasmussen2006}. This can be formally written as
\begin{equation}
	f(x) \sim \mathcal{GP}(m(x), k(x,x')),
\end{equation}
where $m(x) = \mathbb{E}[f(x)]$ is the mean function and $k(x,x') = \mathbb{E}[(f(x)-m(x))(f(x')-m(x'))]$ is the covariance function (or kernel), which describes the correlation between the function values at different inputs.

In practice, the observed data are modeled as $y = f(x) + \epsilon$, with the noise term $\epsilon$ typically assumed to be Gaussian, i.e., $\epsilon \sim \mathcal{N}(0, \sigma_n^2)$.

Given the training data $(X, y)$ and the test inputs $X_*$, the joint distribution over the training outputs and the function values at the test inputs is
\begin{equation}
\begin{pmatrix}
	y \\
	f_*
\end{pmatrix}
\sim \mathcal{N}\left(0, \begin{pmatrix}
	K(X,X) + \sigma_n^2 I & K(X, X_*) \\
	K(X_*, X) & K(X_*, X_*)
\end{pmatrix}\right),
\end{equation}
where $K(X,X)$ is the covariance matrix evaluated at the training inputs, $K(X, X_*)$ is the covariance between the training and test inputs, and $K(X_*, X_*)$ is the covariance matrix at the test inputs.

Using the conditional properties of multivariate Gaussian distributions, the predictive distribution for $f_*$ given the training data and test inputs is
\begin{equation}
f_* \mid X, y, X_* \sim \mathcal{N}(\bar{f}_*, \operatorname{cov}(f_*)),
\end{equation}
where
\begin{equation}
\bar{f}_* = K(X_*, X)\left[K(X,X) + \sigma_n^2 I\right]^{-1} y,
\end{equation}
and
\begin{equation}
\operatorname{cov}(f_*) = K(X_*, X_*) - K(X_*, X)\left[K(X,X) + \sigma_n^2 I\right]^{-1}K(X, X_*).
\end{equation}

This formulation not only yields predictions for the target function values but also provides principled estimates of the associated uncertainties, making GPR a valuable tool in a wide range of data analysis tasks.

Despite its advantages, traditional GPR faces two main issues:
\begin{enumerate}
	\item \textbf{Trend Mismatch:} When the data exhibits a strong global trend, assuming a zero-mean prior without appropriate detrending forces the kernel to model both the trend and local fluctuations. This can reduce the kernel’s flexibility, limit its capacity to capture fine-scale structure, and increase the risk of overfitting \citep{Rasmussen2006}.
	\item \textbf{Kernel Selection:} The choice of covariance function is manual and critical. Different kernel choices imply different prior assumptions, which can lead to substantially different inferences \citep{Rasmussen2006,johnson2025,Zhang2023}.
\end{enumerate}

\subsubsection{Artificial Neural Networks}

As universal approximators, neural networks are extensively employed to model complex functions through pattern learning from data \citep{Cybenko1989}. Mathematically, an ANN consists of layers of interconnected neurons, where each neuron performs a weighted sum of its inputs followed by a nonlinear activation function \citep{Goodfellow2016}. The function approximation performed by an ANN can be expressed as:
\begin{equation}
	y = f_{\text{ANN}}(x; \theta),
\end{equation}
where $ x $ denotes the input data, $ y $ is the corresponding output, and $ \theta $ comprises the set of trainable parameters, including weights and biases.

In practice, a feedforward ANN is typically structured as a sequence of layers: starting with an input layer, followed by one or more hidden layers, and culminating in an output layer. The transformation from layer to layer can be described recursively as:
\begin{equation}
	h^{(l)} = \sigma(W^{(l)} h^{(l-1)} + b^{(l)}),
\end{equation}
where $h^{(l)}$ represents the activation vector at the $l$-th layer, and $ W^{(l)} $, $ b^{(l)} $ are the associated weight matrix and bias vector. The function $ \sigma(\cdot) $ denotes a nonlinear activation function, such as the ReLU ($\max(0, x)$) or the sigmoid function ($1/(1 + e^{-x})$).

Training an ANN involves optimizing the parameters $\theta$ using a dataset $\{(x_i, y_i)\},i=1,...,N,$ to minimize a predefined loss function. This optimization is performed using gradient-based methods such as stochastic gradient descent (SGD) or its variants (e.g., Adam, RMSprop).

ANNs have been widely applied in the astronomical domain due to their flexibility and ability to model complex, nonlinear relationships \citep{Wang2020,Huang2025}.  However, ANNs cannot naturally account for the correlations between data points, a feature that is particularly important in astronomical observations.  Moreover, compared to GPR, the error estimates produced by ANNs lack interpretability.

To address the first limitation discussed above, some researchers have proposed modifying the loss function to
\begin{equation}
	\mathcal{L} = \Delta y^T K^{-1} \Delta y,
\end{equation}
where $\Delta y = y - f_{\text{ANN}}(x; \theta)$ and $K$ is the covariance matrix of the observed data \citep{Dialektopoulos2023}. This loss function is proportional to the negative log-likelihood of a Gaussian distribution (omitting constant terms), effectively incorporating the correlations between data points. In this framework, the observational data are assumed to follow a Gaussian process characterized by the covariance matrix $K$, while the neural network is employed to learn the mean function of this process.

In this work, we take a step further by introducing a new framework that employs two independently trained neural networks: one to model the mean function, and another to represent the kernel (i.e., covariance structure). This separation allows for greater modeling flexibility and enables Bayesian inference within the Gaussian process framework. By directly learning both components from data, the approach improves not only the expressiveness of the model but also the transparency of the resulting uncertainty estimates, which is essential when interpreting astronomical datasets.

\subsubsection{Neural Kernel Gaussian Process Regression}
\paragraph{Mean Net}
In this work, we aim to reconstruct SNe Ia data, where the apparent magnitude $m_{\mathrm{B}}$ exhibits a significant increasing trend with redshift in the low-redshift regime. To avoid the limitations of a zero-mean prior in GPR, we use a neural network to learn a data-driven mean function. 
Due to the scarcity of high-redshift data, directly learning the mean function may lead to poor extrapolation in that region. To address this, we introduce data augmentation by selecting the highest $h$ redshift points and generating $ N_{\text{aug}} $ simulated data points for each. Rather than duplicating the data, we generate each augmented point by sampling from a Gaussian distribution centered on the observed value, with the standard deviation equal to the corresponding measurement error. This avoids introducing excessive redundancy or artificial correlations.

\begin{figure*}
	\includegraphics[width=2\columnwidth]{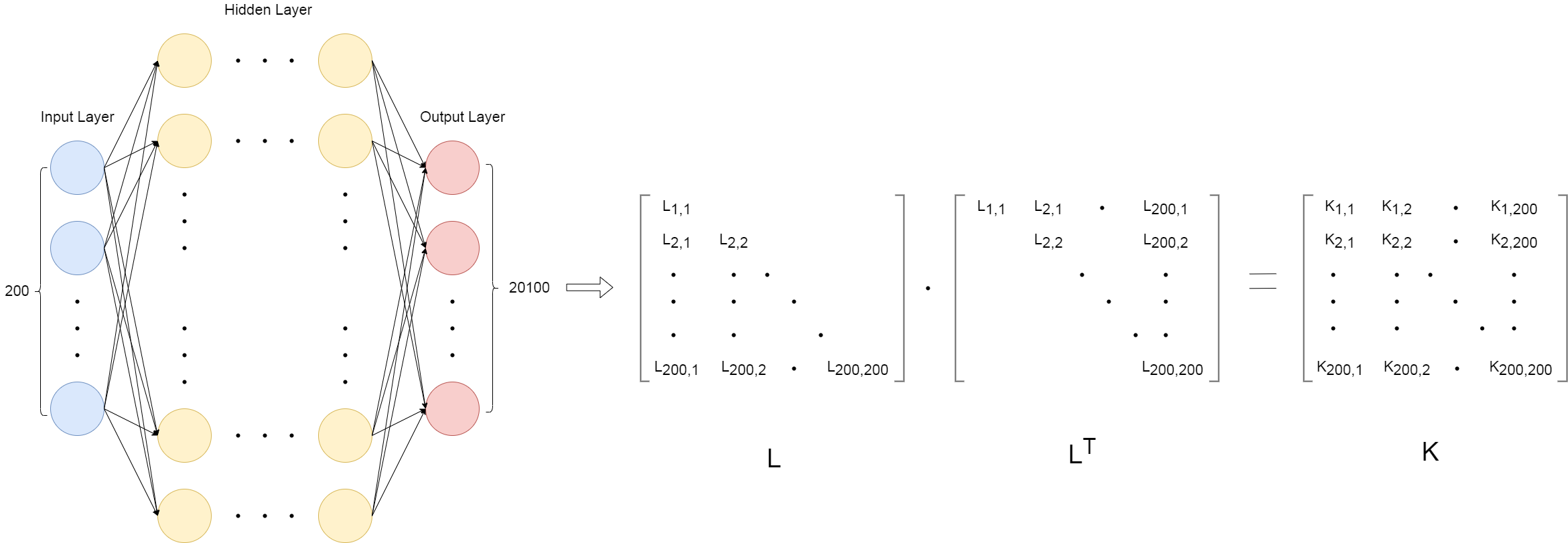}
	\caption{Schematic of the Kernel Net architecture used to learn the covariance matrix $K$. Left: The neural network structure, with an input layer of 200 neurons (representing the feature vector), hidden layers, and an output layer of 20100 neurons (corresponding to the elements of a lower triangular matrix). Right: The output is reshaped into a $200 \times 200$ lower triangular matrix $L$, which is then multiplied by its transpose to form the initial covariance matrix $K = L L^T$, ensuring positive semi-definiteness via Cholesky decomposition. The final covariance matrix is adjusted by element-wise multiplication with a symmetric scaling matrix $S$, where $S_{i,j} = S_{j,i} = \frac{1}{\sqrt{i + j + 1}}$, to mitigate structural bias.}
	\label{fig:Kernel Network-structure}
\end{figure*}

\paragraph{Kernel Net}  
To address the issue of manually selecting a covariance function in GPR, we construct another neural network to learn the kernel function. To ensure that the learned kernel function is symmetric and positive semi-definite, we directly learn the covariance matrix of a set of feature vectors. This is achieved through the Cholesky decomposition, which guarantees positive semi-definiteness. Specifically, the network takes a set of vectors of length 200 as input and outputs a vector of size $200 \times (200+1)/2$. This output is then reshaped into a lower triangular matrix, which is multiplied by its transpose to form the final covariance matrix $K$:
\begin{equation}
	K = LL^T,
\end{equation}
where $L$ is the learned lower triangular matrix. We show the structure of our neural network in Fig.~\ref{fig:Kernel Network-structure}.

However, this approach tends to produce structurally biased covariance matrices. Specifically, the top-left elements (e.g., $K_{11} = L_{11}^2$) involve fewer terms than bottom-right elements (e.g., $K_{nn} = \sum_{k=1}^{n} L_{nk}^2$), leading to a scale imbalance. To mitigate this, we apply a predefined scaling matrix $S$ to balance the structure, where the elements of $S$ are defined as $S_{i,j} = \frac{1}{\sqrt{i + j + 1}}$ for $i, j = 1, \ldots, n$. The final covariance matrix is then computed as $K' = K \odot S$, where $\odot$ denotes the Hadamard (element-wise) product. This approach effectively flattens the covariance matrix scale while preserving its positive semi-definiteness and symmetry \footnote{By the Schur Product Theorem, the Hadamard product of two positive semi-definite matrices is also positive semi-definite. Since $S$ is symmetric and has been verified to be positive semi-definite, $K' = K \odot S$ remains positive semi-definite}.

\paragraph{Training Objective:}  

The observational dataset is split into two parts with a 15:1 ratio. The first subset is used to learn the prior distributions of the mean and kernel functions, while the second part serves as the conditioning set for Bayesian inference, facilitating the computation of the posterior distribution for prediction points.
The first part is further divided into training and validation sets at a 4:1 ratio, enabling effective model training and performance evaluation.

We train the Mean Net by minimizing the mean squared error (MSE) between the predicted and observed apparent magnitudes. The loss function is given by:
\begin{equation}
	\mathcal{L}_{\text{Mean}} = \frac{1}{N} \sum_{i=1}^N \left( \hat{m}_{\rm B}^{(i)} - m_{\mathrm{B}}^{(i)} \right)^2,
\end{equation}
where $\hat{m}_{\rm B}^{(i)}$ is the predicted apparent magnitude, $m_{\mathrm{B}}^{(i)}$ is the observed magnitude, and $N$ is the total number of data points.

To ensure consistency between the learned covariance structure and the uncertainties inferred from data, we minimize the LogDet divergence between the predicted covariance matrix $K$ and the observational covariance matrix $K_{\text{obs}}$. The divergence is defined as:
\begin{equation}
	D_{\text{LD}}(K, K_{\text{obs}}) = \text{tr}(K^{-1} K_{\text{obs}}) + \log \det K - \log \det K_{\text{obs}} - n,
\end{equation}
where $\text{tr}(\cdot)$ denotes the matrix trace, $\det(\cdot)$ denotes the determinant, and $n = 200$ is the dimension of the matrices.

As a Bregman matrix divergence, the LogDet divergence accounts for both spectral misalignment and volumetric distortion between two positive definite matrices. These characteristics make it well-suited for probabilistic modeling, especially in GPR, where accurately capturing uncertainty and structure is essential.

We employ the ReLU activation function in both networks, chosen for its computational efficiency and its capacity to alleviate the vanishing gradient issue. Hyperparameter tuning is performed via Optuna, an automated optimization framework. For the Mean Net, Optuna optimizes the number of layers (ranging from 1 to 3), units per layer (64 to 512 in steps of 32), high-redshift count (1 to 10), and augmentation multiplier (selected from $\{5, 10, 15, 20, 25, 30\}$). For the Kernel Net, the number of layers (1 to 3) and units per layer (chosen from $\{128, 256, 512, 1024, 2048, 4096, 8192\}$) are similarly tuned. The training process leverages a dynamic learning rate schedule and early stopping, where the validation loss is monitored to guide training: the learning rate is reduced if the validation loss does not decrease over a predefined number of epochs, and training terminates early if no improvement is observed within a set number of epochs. This adaptive strategy enhances training stability, accelerates convergence, prevents overfitting, and reduces computational cost. 
The optimal hyperparameter configurations for both networks are summarized in Tab~\ref{tab:hyperparams}.

\begin{table}
	\centering
	\caption{Optimal Hyperparameters for Mean Net and Kernel Net}
	\label{tab:hyperparams}
	\begin{tabular}{lcc}
		\toprule
		\textbf{Hyperparameter} & \textbf{Mean Net} & \textbf{Kernel Net} \\
		\midrule
		Number of layers & 3 & 1 \\
		Units in layer 0 & 256 & 256 \\
		Units in layer 1 & 480 & -- \\
		Units in layer 2 & 512 & -- \\
		High-redshift count ($h$) & 9 & -- \\
		Augmentation multiplier ($N_{\text{aug}}$) & 25 & -- \\
		\bottomrule
	\end{tabular}
	\footnotetext{Values for Mean Net correspond to the neural network used to learn the mean function of the apparent magnitude $m_B$. Values for Kernel Net correspond to the neural network used to learn the covariance structure. Dashes (--) indicate hyperparameters not applicable to the respective network.}
\end{table}

By integrating the learned mean and covariance functions, we construct a fully data-driven Gaussian Process model. Compared to standard neural networks, this approach enhances both interpretability and accuracy in reconstructing supernova data.

\paragraph{Prediction with NKGPR}  
To predict the apparent magnitude $m_{\mathrm{B}}$ at a new redshift point $z_{\text{new}}$, we follow a structured prediction procedure based on the trained Mean Net and Kernel Net. First, the Mean Net is used to predict the prior mean for $z_{\text{new}}$. Next, we combine $z_{\text{new}}$ with the conditioning set (the second part of the dataset split at a 15:1 ratio) to form a vector of length 200, which is input into the Kernel Net. The Kernel Net outputs the complete covariance matrix for this combined set, from which we extract the submatrices $\Sigma_{11}$ (covariance of the conditioning set), $\Sigma_{12}$ (cross-covariance between the conditioning set and $z_{\text{new}}$), and $\Sigma_{22}$ (covariance at $z_{\text{new}}$). Using these, we apply Bayesian inference to compute the posterior distribution of $m_{\mathrm{B}}$ at $z_{\text{new}}$ via the conditional Gaussian distribution formula:
\begin{equation}
	m_{\mathrm{B}}(z_{\text{new}}) \mid \text{data} \sim \mathcal{N}\left( \mu_{\text{post}}, \Sigma_{\text{post}} \right),
\end{equation}
where
\begin{equation}
	\mu_{\text{post}} = \mu(z_{\text{new}}) + \Sigma_{12}^T \Sigma_{11}^{-1} (m_{\mathrm{B}}^{\text{cond}} - \mu^{\text{cond}}),
\end{equation}
\begin{equation}
	\Sigma_{\text{post}} = \Sigma_{22} - \Sigma_{12}^T \Sigma_{11}^{-1} \Sigma_{12},
\end{equation}
with $\mu(z_{\text{new}})$ being the prior mean from the Mean Net, $m_{\mathrm{B}}^{\text{cond}}$ being the observed magnitudes in the conditioning set, and $\mu^{\text{cond}}$ being the prior means for the conditioning set. This method ensures that the prediction incorporates both the learned mean and covariance structures, providing a robust and interpretable estimate of $m_{\mathrm{B}}$ at new redshifts.

\subsection{Testing CDDR with parameterized models}

To investigate potential deviations from the cosmic distance duality relation (CDDR), we introduce a phenomenological function $\eta(z)$ defined as
\begin{equation}
	\eta(z) \equiv \frac{D_{\mathrm{L}}}{D_{\mathrm{A}}} (1+z)^{-2}.
\end{equation}

We consider three commonly used parameterizations of $\eta(z)$ that allow for redshift-dependent deviations:
\begin{itemize}
	\item[P1:] $\eta(z) = 1 + \eta_0 z$ \citep{Bassett2004};
	\item[P2:] $\eta(z) = 1 + \eta_0 \frac{z}{1+z}$ \citep{Holanda2010};
	\item[P3:] $\eta(z) = 1 + \eta_0 \ln(1+z)$ \citep{Nair2011}.
\end{itemize}

In these models, the standard CDDR corresponds to $\eta_0 \equiv 0$. Any statistically significant deviation from this baseline could be indicative of exotic physics. Accordingly, we aim to constrain $\eta_0$ in order to test the validity of the CDDR.

We begin with the standard form of the chi-squared statistic, constructed from the deviation of the theoretical prediction $\eta(z; \eta_0)$ from the observational estimates $\eta_{\rm obs}(z_i)$:
\begin{equation}
	\chi^{2}(\eta_0)=\sum_{i=1}^{N}\frac{\left[\eta(z_i; \eta_0)-\eta_{{\rm obs},i}\right]^2}{\sigma^2_{\eta_{{\rm obs},i}}},
\end{equation}
where $\sigma_{\eta_{\rm obs},i}$ denotes the total uncertainty associated with the observational estimate of $\eta(z_i)$.

To avoid the influence of different prior assumptions on $M_{\mathrm{B}}$ and $r_{\mathrm{d}}$ on our results, we combine them into a single nuisance parameter $\kappa \equiv 10^{M_{\mathrm{B}}/5} r_{\mathrm{d}}$, following \citet{Conley2010,Wang2024}. 

Assuming a flat prior for $\kappa$, we analytically marginalize over it using the formalism of \citet{Wang2024}. The resulting marginalized chi-squared statistic is:
\begin{equation}
	\chi_{\rm M}^{\prime\,2}(\eta_0) = C - \frac{B^2}{A} + \ln\left( \frac{A}{2\pi} \right),
\end{equation}
where: $A = \sum_{i=1}^{N} \frac{\eta(z_i; \eta_0)^2}{\sigma_{\eta_{\rm obs},i}^{\prime\,2} \beta_i^2}$, $B = \sum_{i=1}^{N} \frac{\eta(z_i; \eta_0)}{\sigma_{\eta_{\rm obs},i}^{\prime\,2} \beta_i}$, $C = \sum_{i=1}^{N} \frac{1}{\sigma_{\eta_{\rm obs},i}^{\prime\,2}}$, with $\beta_i = 10^{(m_{{\rm B},i}/5 - 5)} \theta_{{\rm BAO},i}(1 + z_i)^{-1}$ and the variance defined as:
\begin{equation}
	\sigma_{\eta_{\rm obs},i}^{\prime\,2} = \left( \frac{\ln 10}{5} \, \sigma_{m_{{\rm B},i}} \right)^2 + \left( \frac{\sigma_{\theta_{{\rm BAO},i}}}{\theta_{{\rm BAO},i}} \right)^2.
\end{equation}

This marginalized statistic $\chi_{\rm M}^{\prime\,2}(\eta_0)$ is completely independent of $M_{\mathrm{B}}$, $r_{\mathrm{d}}$, and consequently also of the Hubble constant $H_0$. We determine the best-fit value of $\eta_0$ by minimizing $\chi_{\rm M}^{\prime\,2}$, and assess the constraints on each parameterization using the corresponding likelihood function.

\section[RESULTS AND ANALYSIS]{RESULTS AND ANALYSIS}\label{sec:RESULTS AND ANALYSIS}

In this section, we present the outcomes of our analysis using the NKGPR method to reconstruct the supernova apparent magnitude $ m_{\mathrm{B}} $ and test the CDDR. The results are divided into two main parts: the reconstruction of $ m_{\mathrm{B}} $ from the Pantheon+ supernova data and the subsequent CDDR test using DESI BAO data with three parameterized forms of $\eta(z)$. Finally, we use the AIC and BIC to compare the results obtained from the three parameterizations.

\subsection{Reconstruction of Supernova Apparent Magnitude $ m_{\mathrm{B}} $}

The NKGPR method leverages two neural networks, Mean Net and Kernel Net, to learn the mean function and covariance structure of the Pantheon+ supernova data. Below, we detail the reconstruction process and its outcomes.

\subsubsection{Mean Function Reconstruction}

\begin{figure}
	\includegraphics[width=\columnwidth]{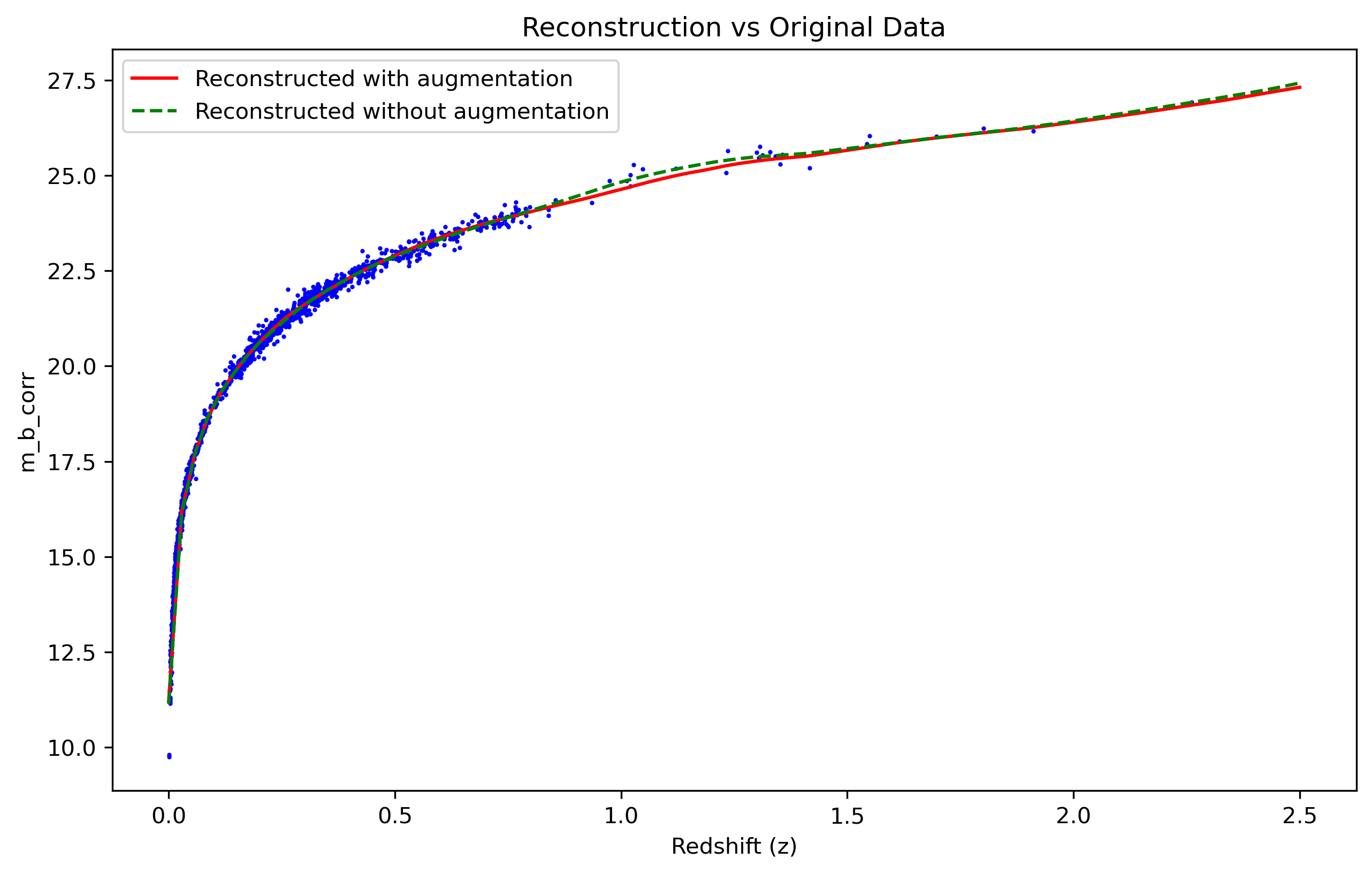}
	\caption{Reconstruction of the apparent magnitude \( m_{\mathrm{B}} \) using Pantheon+ supernova data with the Mean Net including data augmentation, and without augmentation for comparison.}
	\label{fig:mean_reconstruction}
\end{figure}

The Mean Net reconstructs the mean function of the apparent magnitude $ m_{\mathrm{B}} $ as a function of redshift $ z $. Figure \ref{fig:mean_reconstruction} illustrates this result, displaying the original Pantheon+ supernova data as scatter points alongside the reconstructed mean function as a solid curve. The plot demonstrates that the Mean Net successfully captures the increasing trend of $ m_{\mathrm{B}} $ at low redshifts ($ z < 1 $) and provides a reliable extrapolation at higher redshifts ($ z > 1 $), despite the sparsity of data in that regime. This smooth reconstruction validates the effectiveness of the neural network in modeling the global behavior of the supernova data. For comparison, we also show reconstruction of the apparent magnitude \( m_{\mathrm{B}} \) using Pantheon+ supernova data without augmentation in Figure \ref{fig:mean_reconstruction}.

\subsubsection{Covariance Matrix Reconstruction}

\begin{figure}
	\includegraphics[width=\columnwidth]{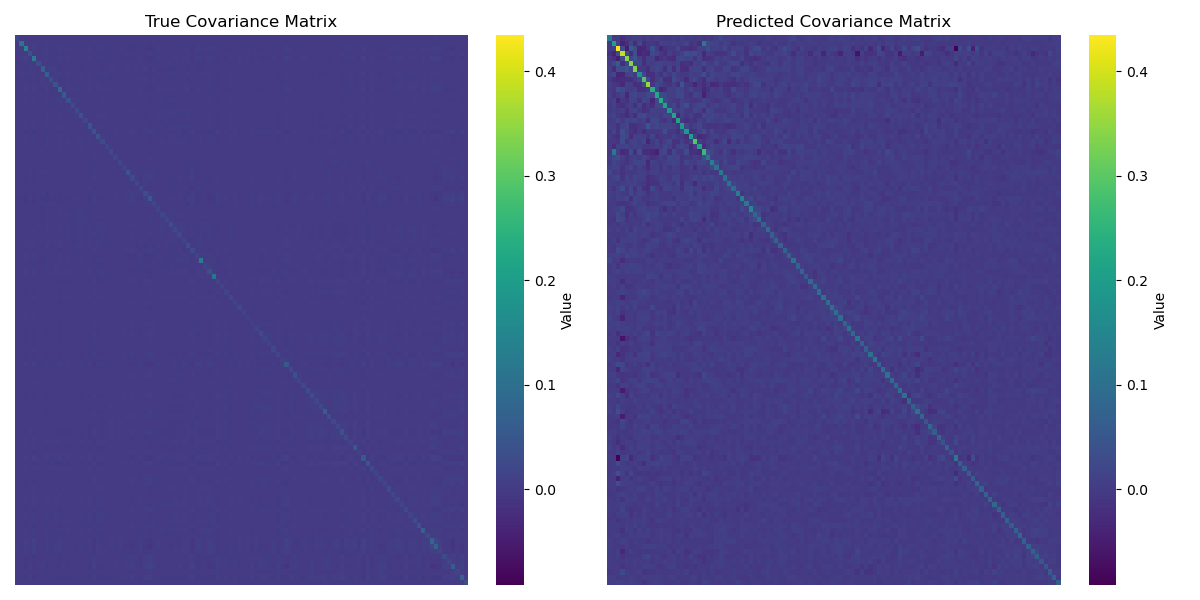}
	\caption{Comparison of observed and predicted covariance matrices. \textit{Left}: Heatmap of the observed covariance matrix $ \Sigma_{11} $. \textit{Right}: Heatmap of the predicted covariance matrix $ \Sigma_{11}^{\text{pre}} $ learned by the Kernel Net.}
	\label{fig:covariance_comparison}
\end{figure}

To evaluate the Kernel Net’s ability to learn the covariance structure, we compare the observed covariance matrix $ \sigma_{11} $ (derived from the Pantheon+ data) with the predicted covariance matrix $ \sigma_{11}^{\text{pre}} $. Figure \ref{fig:covariance_comparison} presents this comparison through two heatmaps: the left panel shows $ \sigma_{11} $, and the right panel shows $ \sigma_{11}^{\text{pre}} $. The visual similarity between the two matrices suggests that the Kernel Net accurately reproduces the correlation patterns in the observational data. This fidelity is critical for ensuring reliable uncertainty estimates in the Gaussian process framework, particularly for the subsequent Bayesian inference step.

\subsubsection{Final Prediction of $ m_{\mathrm{B}} $ with Uncertainty}

\begin{figure}
	\centering
	\begin{subfigure}{\columnwidth}
		\includegraphics[width=\textwidth]{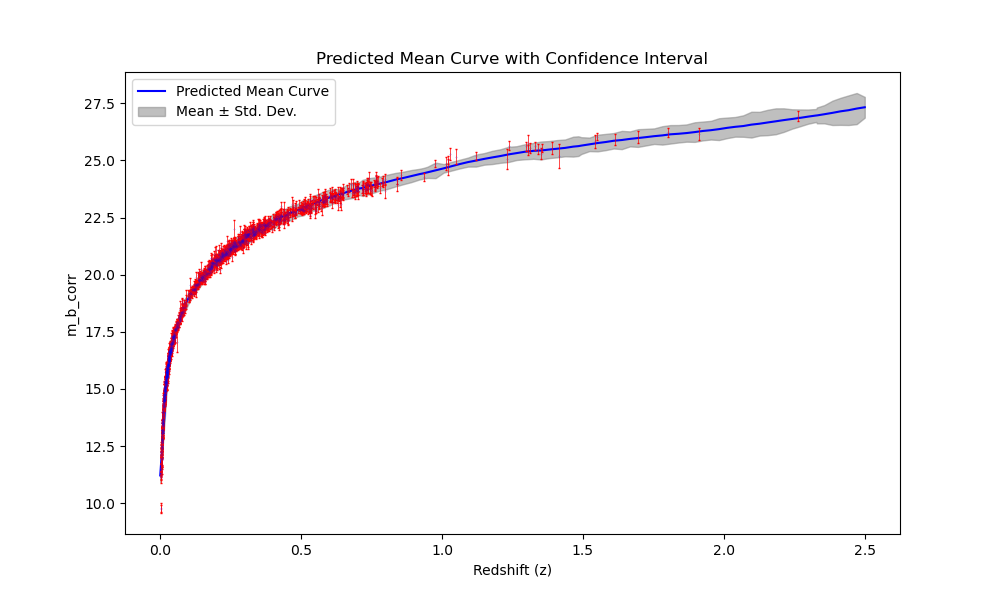}
		\caption{NKGPR}
		\label{fig:predict_m_b_nkgpr}
	\end{subfigure}
	\\[1ex] 
	\begin{subfigure}{\columnwidth}
		\includegraphics[width=\textwidth]{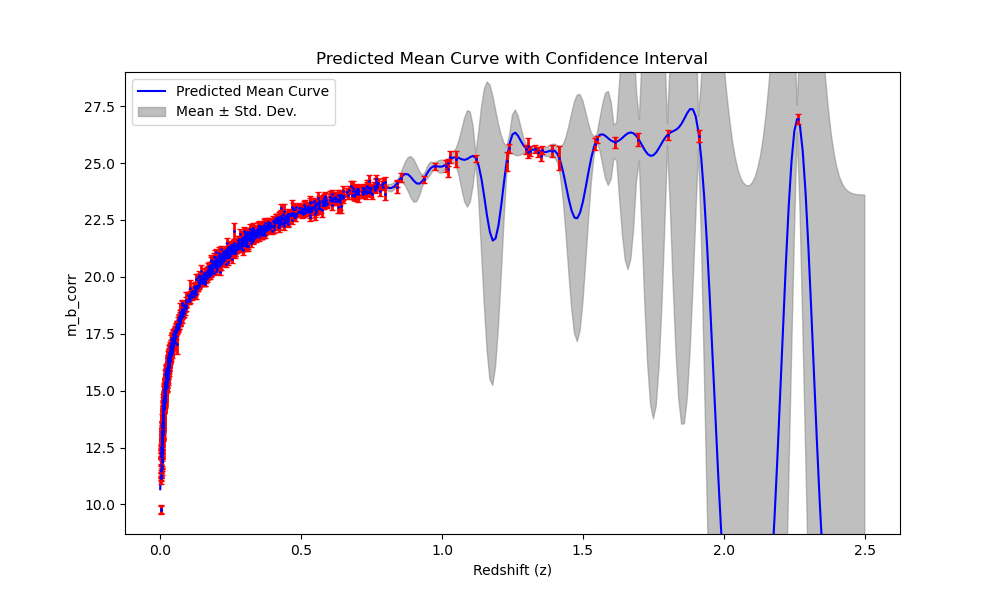}
		\caption{GaPP}
		\label{fig:predict_m_b_gapp}
	\end{subfigure}
	\caption{Comparison of the final prediction of the apparent magnitude $m_{\mathrm{B}}$ using NKGPR (top) and GaPP (bottom). The solid lines represent the predicted means, and the shaded regions denote the 1$\sigma$ uncertainty bands.}
	\label{fig:predict_m_b_comparison}
\end{figure}

Using the mean function from the Mean Net and the covariance matrix from the Kernel Net, we perform Bayesian inference to derive the final prediction of $ m_{\mathrm{B}} $ and its associated uncertainty. Figure \ref{fig:predict_m_b_comparison} compares the predicted mean curve of $ m_{\mathrm{B}} $ as a function of redshift $ z $ using our NKGPR method and the traditional GaPP method. The solid lines represent the predicted means, and the shaded regions denote the 1$\sigma$ uncertainty bands.

The NKGPR method (top panel) produces a smoother predicted curve that aligns well with the observed data points, with uncertainty bands that widen gradually at higher redshifts due to the limited number of data points. In contrast, the GaPP method with an RBF kernel (bottom panel), exhibits more oscillatory behavior in the predicted mean, particularly at $ z > 1.0 $, and significantly wider uncertainty bands. This behavior in GaPP arises because it adopts a simple zero-mean prior, which fails to capture the significant upward trend in the apparent magnitude $ m_{\mathrm{B}} $ data. As a result, the covariance function in GaPP compensates for this global trend, leading to overfitting, oscillations, and inflated uncertainties, especially in regions with sparse data. The NKGPR method provides a more robust estimate, offering a better balance between accuracy and error quantification in such sparse regions. This reconstructed $ m_{\mathrm{B}} $ serves as the foundation for computing LD in the CDDR test.

\subsection{Testing the Cosmic Distance Duality Relation}

With the reconstructed $ m_{\mathrm{B}} $ and the corresponding LD, we test the CDDR by combining these with ADD derived from DESI BAO data. We adopt three parameterized forms for the deviation function $\eta(z)$:

\begin{itemize}
	\item[P1:] $\eta(z) = 1 + \eta_0 z$
	\item[P2:] $\eta(z) = 1 + \eta_0 \frac{z}{1+z}$
	\item[P3:] $\eta(z) = 1 + \eta_0 \ln(1+z)$
\end{itemize}

For each model, we constrain the parameter $\eta_0$ using the marginalized $\chi^2$ statistic described in Section \ref{sec:DATA AND METHODOLOGY}.

\subsubsection{Likelihood Functions for $\eta_0$}

\begin{figure}
	\includegraphics[width=\columnwidth]{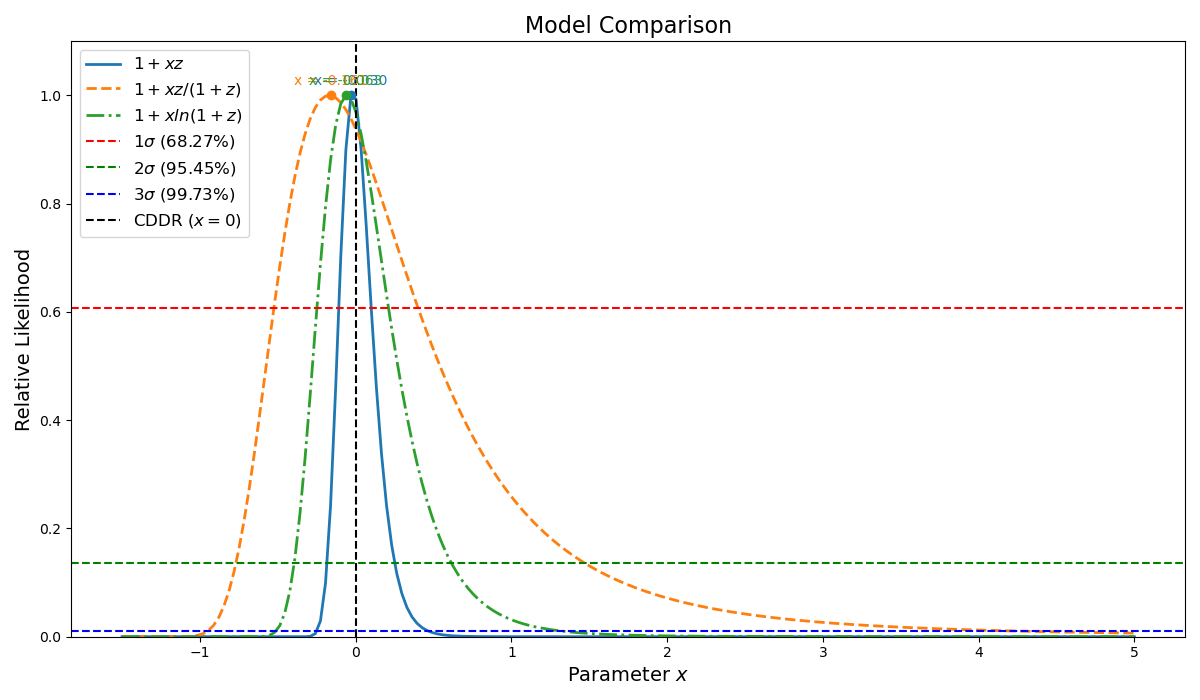}
	\caption{Relative likelihood functions for $\eta_0$ in the three parameterizations of $\eta(z)$: P1 (solid line), P2 (dashed line), and P3 (dotted line).}
	\label{fig:likelihoods}
\end{figure}

Figure \ref{fig:likelihoods} plots the relative likelihood functions for $\eta_0$ across the three parameterizations on a single graph. The curves are normalized to their peak values, representing the maximum likelihood estimates. All three likelihoods peak near $\eta_0 = 0$, suggesting consistency with the standard CDDR. However, the shapes and widths of the likelihoods differ, reflecting varying degrees of constraint imposed by each parameterization on the data.

\subsubsection{Maximum Likelihood Results and Confidence Intervals}

Table \ref{tab:eta0_constraints} lists the best-fit values of $\eta_0$ for each parameterization, along with their 1$\sigma$, 2$\sigma$, and 3$\sigma$ confidence intervals. The results show that $\eta_0$ is consistent with zero within the 1$\sigma$ interval for all models, indicating no significant deviation from the CDDR. To assess the reliability of data augmentation, we conducted a cross-validation using the Pantheon+ supernova data without augmentation. The reconstruction is shown in Figure~\ref{fig:mean_reconstruction}, and the resulting maximum likelihood estimates (MLE) and 1$\sigma$ confidence intervals for the models are: P1 (\(1 + xz\)): MLE = 0.003, 1$\sigma$ = (-0.063, 0.101); P2 (\(1 + xz/(1+z)\)): MLE = -0.030, 1$\sigma$ = (-0.422, 0.590); P3 (\(1 + x\ln(1+z)\)): MLE = 0.003, 1$\sigma$ = (-0.193, 0.264). These results support the robustness of our \(\eta_0\) constraints.
The P2 model exhibits the broadest confidence intervals, suggesting it imposes the weakest constraint, while P1 and P3 provide tighter bounds.

\begin{table}
	\centering
	\caption{Best-fit values of $\eta_0$ and confidence intervals for the three parameterizations.}
	\label{tab:eta0_constraints}
	\setlength{\tabcolsep}{3pt}
	\begin{tabular}{lcccc}
		\toprule
		\textbf{Model} & \textbf{ $\eta_0$} & \textbf{1$\sigma$ Interval} & \textbf{2$\sigma$ Interval} & \textbf{3$\sigma$ Interval} \\
		\midrule
		P1: $1 + \eta_0 z$ & -0.030 & (-0.095, 0.068) & (-0.161, 0.231) & (-0.226, 0.460) \\
		P2: $1 + \eta_0 \frac{z}{1+z}$ & -0.161 & (-0.520, 0.394) & (-0.749, 1.472) & (-0.945, 4.151) \\
		P3: $1 + \eta_0 \ln(1+z)$ & -0.063 & (-0.226, 0.198) & (-0.389, 0.590) & (-0.487, 1.276) \\
		\bottomrule
	\end{tabular}
\end{table}

\subsection{Model Comparison Using AIC and BIC}

The Akaike Information Criterion (AIC) and Bayesian Information Criterion (BIC) are widely used metrics for model selection in statistical analysis. They balance the goodness of fit, typically measured by the log-likelihood (or equivalently, the minimum $\chi^2_{\text{min}}$ in our context), with a penalty for model complexity to prevent overfitting. Specifically, AIC is defined as $\text{AIC} = \chi^2_{\text{min}} + 2k$, where $k$ is the number of model parameters, while BIC is given by $\text{BIC} = \chi^2_{\text{min}} + k \ln N$, with $N$ being the number of data points. The penalty term in AIC focuses on predictive accuracy, whereas BIC imposes a stronger penalty for complexity, favoring the true model when $N$ is large. These criteria help identify the model that best explains the data with the least complexity, making them essential tools for comparing competing models.

In this study, since each of the three parameterizations considered in this study involve only one free parameter, the penalty terms in AIC and BIC are identical, making these information criteria effectively equivalent to a direct comparison of the minimum $\chi^2$ values.
Therefore, computing the AIC and BIC is not essential in this particular analysis. Nevertheless, we report them for completeness and to enable consistent comparisons with more complex parameterizations in future work as well as with related studies in the literature.

Table \ref{tab:aic_bic} presents these values, along with the differences $\Delta$AIC and $\Delta$BIC relative to the model with the lowest AIC and BIC (P2 in both cases). 
\footnote{Note that the value of $N$ in the BIC calculation is the data pairs of SNe Ia and BAOs which corresponds to the number of DESI BAO data points ($N = 6$), as the AIC and BIC are computed solely from the likelihood analysis stage, independently of the earlier data reconstruction process.}
Among the three parameterizations, P2 achieves the lowest AIC (13.947) and BIC (13.556), making it the statistically preferred model under both criteria. Nevertheless, the best-fit results for all three models are consistent with $\eta_0 = 0$ within the 1$\sigma$ confidence interval, and the small $\Delta$AIC and $\Delta$BIC values for P1 and P3 (less than 0.15) suggest that there is no significant deviation from the CDDR and no clear statistical preference among the models.

\begin{table}
	\centering
	\caption{AIC and BIC values for the three parameterizations, with $\Delta$AIC and $\Delta$BIC relative to the best model (P2).}
	\label{tab:aic_bic}
	\begin{tabular}{lcccc}
		\toprule
		\textbf{Model} & \textbf{AIC} & \textbf{$\Delta$AIC} & \textbf{BIC} & \textbf{$\Delta$BIC} \\
		\midrule
		P1: $1 + \eta_0 z$ & 13.944 & 0.121 & 13.736 & 0.121 \\
		P2: $1 + \eta_0 \frac{z}{1+z}$ & 13.823 & 0.000 & 13.615 & 0.000 \\
		P3: $1 + \eta_0 \ln(1+z)$ & 13.891 & 0.068 & 13.683 & 0.068 \\
		\bottomrule
	\end{tabular}
\end{table}

\section{Conclusion}
\label{sec:CONCLUSION}

In this work, we introduce a new method, Neural Kernel Gaussian Process Regression, to test the CDDR. First, we use NKGPR to reconstruct the apparent magnitude $m_{\mathrm{B}}$ from Pantheon+ data to align with DESI data redshifts. Then, to ensure a model-independent approach, we marginalize over the nuisance parameters $M_{\mathrm{B}}$ and $r_{\mathrm{d}}$. Finally, we test three parameterizations of $\eta(z)$ to probe deviations from $\eta(z) = 1$.

Our results show that the reconstructed supernova magnitudes are consistent with the Pantheon+ data at low redshift and provide reliable extrapolation at high redshift. Compared to traditional methods like GaPP with fixed kernels and zero-mean assumptions, NKGPR exhibits improved smoothness and uncertainty control, particularly in the high-$z$ regime. The subsequent analysis of three widely used parameterizations for $\eta(z)$ yields no statistically significant deviation from the expected CDDR relation $\eta(z) = 1$, indicating strong consistency with current observations.

Nevertheless, several challenges remain.
First, The current implementation fixes the covariance matrix dimension at 200 to balance computational cost and performance. However, this design, together with the $\mathcal{O}(n^3)$ time complexity arising from Cholesky decomposition, limits the scalability and adaptability of NKGPR to larger datasets anticipated in future surveys. A promising direction is to explore low-rank approximations or sparse Gaussian process methods as a means to alleviate the computational burden.
Second, in order to alleviate the sparsity of high-redshift data, we introduced a Gaussian-based data augmentation strategy during the training phase of the Mean Net. While such augmentation may potentially introduce systematic bias due to the use of artificial data, it is important to emphasize that the augmented data are only used in the training set and are excluded from the validation set, and thus do not influence model evaluation. Moreover, we conducted a cross-check in which the Mean Net was trained without augmented data. The resulting cosmological constraints remained statistically consistent with those obtained using the augmented version, demonstrating the robustness of our method. In addition, as future astronomical surveys are expected to provide more high-redshift supernovae, this issue will likely become less relevant in subsequent studies.

In addition, to mitigate structural issues in the predicted covariance matrix, we introduced a scaling matrix $S_{i,j} = \frac{1}{\sqrt{i + j + 1}}$. We verified its positive semi-definiteness and evaluated its performance against alternatives, including the unscaled case and a simpler form $S_{i,j} = \frac{1}{i + j + 1}$. However, this design lacks a formal theoretical derivation, which may limit the robustness of the Kernel Net. A more direct approach would be to replace the generation of a lower triangular matrix with the generation of a full 200×200 matrix and construct the covariance matrix by multiplying it with its own transpose. While this guarantees positive semi-definiteness, it significantly increases computational cost and parameter complexity.

Future work will pursue two key directions. On the methodological front, we aim to develop more flexible and computationally efficient kernel-learning strategies to improve adaptability to the large-scale cosmological datasets. In terms of application, we plan to employ NKGPR in a wider range of astronomical contexts, such as constraining cosmological parameters, including the Hubble constant and the dark energy equation of state, as well as reconstructing the cosmic expansion history. These efforts will further demonstrate the versatility of NKGPR and contribute to a deeper understanding of the universe’s structure and evolution.

In summary, this study validates the CDDR using NKGPR and demonstrates its potential as a versatile tool for cosmological data analysis. We anticipate that NKGPR will contribute significantly to future efforts to explore the universe’s structure and evolution with greater accuracy.

\section*{ACKNOWLEDGEMENTS}
We are grateful to the referee for helpful comments and constructive suggestions. We also thank Zhen Huang and Professor Xiangyun Fu for their valuable assistance during this project. This project was supported by the Guizhou Provincial Science and Technology Foundations (QKHJC-ZK[2024] general 443 and QKHPT ZSYS[2025] 004).

\section*{DATA AVAILABILITY}
Data are available at the following references:
the Pantheon+ SNe Ia sample from \citet{Scolnic2022},
and the DESI Baryon Acoustic Oscillation data from
\citep{DESI:2503.14738, DESI:2503.14739}.



\bibliographystyle{mnras}
\bibliography{mybib} 








\bsp	
\label{lastpage}
\end{document}